\documentclass[a4paper,11pt]{article}
\usepackage{pos}

\newcommand{\Amp}{\mathcal{M}}
\newcommand{\Ham}{\mathcal{H}}

\newcommand{\mx}{\mathcal{F}}
\newcommand{\cng}[1]{\textcolor{black}{#1}}
\title{New physics search via CP observables in $B_s^0 \rightarrow \phi \phi$ decay with Chromomagnetic operators}

\author*[a]{Tejhas Kapoor}

\affiliation[a]{Université Paris-Saclay, CNRS/IN2P3, IJCLab,\\
  91405 Orsay, France}

\emailAdd{tejhas.kapoor@ijclab.in2p3.fr}

\abstract{In this work, we investigate the time-dependent angular analysis of $B_s^0 \rightarrow \phi \phi$ decay to search for new physics signals via CP-violating observables. We work with a new physics Hamiltonian containing both left- and right-handed Chromomagnetic dipole operators. The hierarchy of the helicity amplitudes in this model gives us a new scheme of experimental search, which is different from the ones LHCb has used in its analysis. To illustrate this new scheme, we perform a sensitivity study using two pseudo datasets generated using LHCb's measured values. We find the sensitivity of CP-violating observables to be of the order of $5-7\%$ with the current LHCb statistics. In addition, we present a revised version of the table of coefficients of time-dependent terms in the angular decay distribution with precisely defined quantities.}

\FullConference{%
  *** The European Physical Society Conference on High Energy Physics (EPS-HEP2023), ***\\
  *** 21-25 August 2023 ***\\
  ***  Universität Hamburg, Hamburg, Germany ***
}

\begin{document}
\maketitle

\section{Introduction}
$B_s^0 \rightarrow \phi \phi$ decay is an excellent channel to search for new physics (NP). The presence of penguin quantum loop makes it an excellent probe to search for new heavy particles and being a purely penguin-type decay keeps it free from tree-penguin interference contamination. We choose to search for NP via CP-violating observables as their Standard Model (SM) values are well known: the phase of the indirect CP violation in $B_s^0 \rightarrow \phi \phi$ decay is zero in SM and there is no direct CP violation. Thus, they are clean observables and we can perform a null test on these quantities to search for NP.
\par 
In this work, we present a NP model based on Chromomagnetic operator. This model gives rise to a phase scheme that is different from the one used by LHCb in their work \cite{lhcb}, and a fit using this new scheme would help to search for NP models that can manifest themselves via the Chromomagnetic operator. Finally, based on this new phase scheme, we perform a sensitivity study by generating pseudodata from the LHCb best fit values.
\section{Angular decay distribution} \label{sec:angdecaydist}
\begin{figure}[t]
\setlength{\unitlength}{1mm}
  \centering
  \begin{picture}(140,60)
    \put(0,-1){
      \includegraphics*[width=130mm]{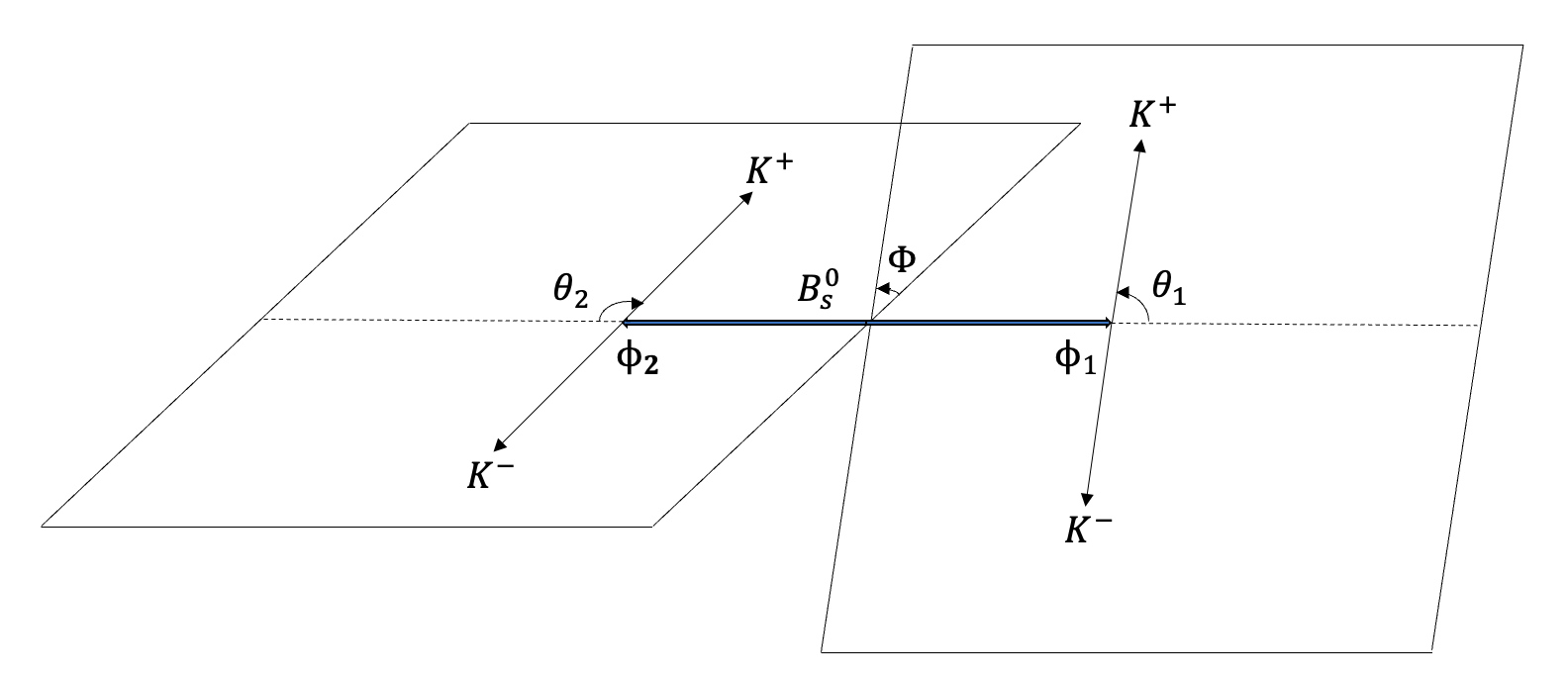}
    }
  \end{picture}
  \caption{\small Decay angles for the {$B_s^0 \rightarrow \phi(\rightarrow K^+K^-) \phi(\rightarrow K^+K^-)$} decay, where  
$\theta_{1(2)}$ is the angle between the $K^+$ momentum in the $\phi_{1(2)}$ meson rest frame and the $\phi_{1(2)}$ momentum in the $B_s^0$ rest frame. $\Phi$ is the angle between the two $\phi$ meson decay planes.}
\label{fig:decayangles}
\end{figure}

The angular decay distribution for {$B^0_s \rightarrow \phi(\rightarrow K^+K^-)\phi(\rightarrow K^+K^-)$} decay can be described by the help of three angles as shown in Figure \ref{fig:decayangles}. The amplitude for this process is given by \cite{lhcb}
\begin{align}\label{amplitude}
\mathcal{A}(t,\theta_1,\theta_2,\Phi) &= A_0(t)\cos\theta_1\cos\theta_2 
+\frac{A_\parallel(t)}{\sqrt{2}}\sin\theta_1\sin\theta_2\cos\Phi 
+i\frac{A_\perp(t)}{\sqrt{2}}\sin\theta_1\sin\theta_2\sin\Phi,
\end{align}
where $A_0$ is the longitudinal CP-even, $A_{\parallel}$ is the transverse-parallel CP-even and $A_{\perp}$ is the transverse-perpendicular CP-odd transversity amplitude. The resulting angular decay distribution is proportional to square of the amplitude in Eq.~\eqref{amplitude}, in which we can separate the time and angular dependence. The time-dependent term $K_i(t)$ is given as ($i$ runs from 1 to 6) \cite{lhcb}
\begin{align} \label{timedep}
\begin{aligned}
K_i(t)=N_i e^{-\Gamma_s t} \bigg[ 
&a_i \cosh\left(\frac{1}{2}\Delta \Gamma_s t\right) + b_i \sinh\left(\frac{1}{2}\Delta\Gamma_s t\right) + 
c_i \cos(\Delta m_s t) + d_i\sin(\Delta m_s t) \bigg].
\end{aligned}
\end{align}
The coefficients $N_i,a_i,b_i,c_i$ and $d_i$ are the LHCb experimental observables given in  Table~\ref{timedeptablestrongCP}. {The structure of these coefficients depend on the form of amplitudes $A_{0,\parallel,\perp}(t)$, defined in Section~\ref{cpvnpparam}}. $\Delta \Gamma_s \equiv \Gamma_L - \Gamma_H$ is decay-width difference between the light and heavy $B_s^0$ mass eigenstate, $\Gamma_s \equiv (\Gamma_L + \Gamma_H)/2$ is the average decay width and $\Delta m_s \equiv m_H - m_L$ is the mass difference between the heavy and light $B_s^0$ mass eigenstate, and also the $B_s^0 - \bar{B}_s^0$ oscillation frequency.

\section{CP violating quantities in the presence of new physics: parametrisation}\label{cpvnpparam}
In this study, we are only probing CP-violating phases in the decay; thus, our parametrisation is done accordingly (the $B_s^0 - \bar{B}_s^0$ mixing amplitude has already been well constrained by previous measurements 
\cite{lhcbjpsimixing}, so we do not focus on it in this work). Also, we assume $|\frac{q}{p}| = 1$ \cite{mixingcpasymmetry}. 
\par
The helicity/transversity amplitudes, with helicity/transversity '$k$' are written as \cite{rosner} 
\begin{align}
\begin{aligned}
A_k(t) &= \langle (\phi\phi)_k | \Ham_{\rm{eff}} | B_s^0(t) \rangle = g_+(t) A_k + \frac{q}{p}g_-(t)\bar{A}_k  \\
\bar{A}_k(t) &= \langle (\phi\phi)_k | \Ham_{\rm{eff}} | \bar{B}_s^0(t) \rangle = g_+(t) \bar{A}_k + \frac{p}{q} g_-(t) A_k.
\end{aligned}
\end{align}
where $g_+(t)$ and $g_-(t)$ describe the time evolution of $B^0_s$ and $\bar{B}^0_s$, respectively.
\cng{The amplitude at $t=0$, comprising of both SM and NP amplitudes} can be written as :
\begin{align} \label{ampNP}
\begin{aligned}
A_k(0) \equiv A_k &= A_k^{\rm SM}+A_k^{\rm NP} \\
&= |A_k^{\rm SM}| e^{i\delta_k^{\rm SM}}e^{i\phi^{\rm SM}} + |A_k^{\rm NP}| e^{i\delta_k^{\rm NP}}e^{i\phi_k^{\rm NP}}  \\
&= |A_k^{\rm SM}| e^{i\delta_k^{\rm SM}}e^{i\phi^{\rm SM}}  \left( 1 + r_k^{\rm NP} e^{i(\phi_k^{\rm NP} - \phi^{\rm SM})} e^{i(\delta_k^{\rm NP} - \delta_k^{\rm SM})} \right) \\
&= |A_k^{\rm SM}| e^{i\delta_k^{\rm SM}}e^{i\phi^{\rm SM}} X_k e^{i\theta_k}, 
\end{aligned}
\end{align}
where $\delta$ is the strong and $\phi$ is the phase. In the last line, we denote the quantity in the parenthesis as $X_k e^{i\theta_k}$ and $r_k^{\rm NP} = \frac{|A_k^{\rm NP}|}{|A_k^{\rm SM}|}$. The phase $\theta_k$ is a \emph{mixture of weak and strong phases}. Similarly for the CP-conjugate amplitude, the expression is ($\eta_k$ is the CP eigenvalue of the transversity state, with $\eta_{\perp} =-1$ and $\eta_{0,\parallel} =1$)
\begin{align}\label{ampNPconj}
\begin{aligned}
\bar{A}_k &= \eta_k |A_k^{\rm SM}| e^{i\delta_k^{\rm SM}}e^{-i\phi^{\rm SM}}  \left( 1 + r_k^{\rm NP} e^{-i(\phi_k^{\rm NP} - \phi^{\rm SM})} e^{i(\delta_k^{\rm NP} - \delta_k^{\rm SM})} \right) \\
&= \eta_k |A_k^{\rm SM}| e^{i\delta_k^{\rm SM}}e^{-i\phi^{\rm SM}} X_k^c e^{i\theta_k^c}.
\end{aligned}
\end{align}
Recalling that $\arg(q/p) = 2\beta_s \approx 2\phi^{\rm SM}$, we finally get 
\begin{align}\label{intphase}
\frac{q}{p}\frac{\bar{A}_k}{A_k} = \eta_k \lambda_k e^{-i(\theta_k - \theta_k^c)},
\end{align}
where $\lambda_k \equiv \frac{|\bar{A}_k|}{|A_k|} = \frac{X^c_k}{X_k}$ becomes the direct CP violation measurement parameter:
$\lambda_k \neq 1$ implies direct CP violation is present in the decay. Since in SM, $\lambda_k =1$ for all helicities, the deviation of this value from $1$ (by more than $O(\lambda^3)$) would be a clear signal for NP, i.e. $\lambda_k-1$ is a null-test parameter for NP.  Another quantity that can be used for NP search is the interference phase $\theta_k-\theta_k^c$. In SM, this quantity is zero; thus, the deviation of this quantity from zero (by more than $O(\lambda^3)$) would be a signal of NP, i.e. $\theta_k-\theta_k^c$ is also a null-test parameter for NP.  
The time-dependent amplitude is given by
\begin{align}\label{timedepamp}
A_k(t) &= A_k \left[ g_+(t) + g_-(t) \frac{q}{p} \frac{\bar{A}_k}{A_k} \right] \\
&=|A_k^{\rm SM}| X_k e^{i\delta_k^{\rm SM}}e^{i\phi^{\rm SM}} e^{i\theta_k} \left[ g_+(t) + g_-(t) \eta_k \lambda_k e^{-i(\theta_k - \theta_k^c)} \right]. 
\end{align} 
The coefficients of the time-dependent terms in Eq.~\eqref{timedep}, obtained by using Eq.~\eqref{timedepamp},  are given in Table~\ref{timedeptablestrongCP}.
\begin{table}[h]
\resizebox{\textwidth}{!}{
  \begin{tabular}{|c|c|c|c|c|c|c|}
  \hline
i    & $N_i$                               & $a_i$                & $b_i$                             & $c_i $              & $d_i$                             & $f_i $ \\ \hline \hline
1    &$ |A_0|^2$                       & ${(1+\lambda_0^2)}/{2}$       & $-\lambda_0 \cos(\theta^c_0 - \theta_0)$                           &  ${(1-\lambda_0^2)}/{2}$            & $-\lambda_0 \sin(\theta^c_0 - \theta_0)$                             &$4\cos^2\theta_1\cos^2\theta_2$ \\ \hline
2    & $|A_\parallel |^2$           &${(1+\lambda_{\parallel}^2)}/{2}$          &$-\lambda_{\parallel} \cos(\theta^c_{\parallel} - \theta_{\parallel})$             & ${(1-\lambda_{\parallel}^2)}/{2}$              &$-\lambda_{\parallel} \sin(\theta^c_{\parallel} - \theta_{\parallel})$                &$\sin^2\theta_1\sin^2\theta_2(1{+}\cos2\Phi)$ \\ \hline
3    &$ |A_\perp |^2 $              &${(1+\lambda_{\perp}^2)}/{2}$       & $\lambda_{\perp} \cos(\theta^c_{\perp} - \theta_{\perp})$                    &${(1-\lambda_{\perp}^2)}/{2}$            & $\lambda_{\perp} \sin(\theta^c_{\perp} - \theta_{\perp})$                &$\sin^2\theta_1\sin^2\theta_2(1{-}\cos2\Phi)$ \\ \hline
4    &$ {|A_\parallel||A_\perp |}/{2}$   
& $\begin{array}{c} \sin(\delta_{\perp} - \delta_{\parallel} + \theta_{\perp} -\theta_{\parallel} )  \\ -\lambda_{\perp} \lambda_{\parallel} \sin(\delta_{\perp} - \delta_{\parallel} + \theta^c_{\perp} - \theta^c_{\parallel} )  \end{array}$
& $\begin{array}{c} \lambda_{\perp} \sin(\delta_{\perp} - \delta_{\parallel} + \theta^c_{\perp} - \theta_{\parallel} ) \\  - \lambda_{\parallel} \sin(\delta_{\perp} - \delta_{\parallel} + \theta_{\perp} - \theta^c_{\parallel} ) \end{array}$
& $\begin{array}{c} \sin(\delta_{\perp} - \delta_{\parallel} + \theta_{\perp} - \theta_{\parallel})  \\ +\lambda_{\perp} \lambda_{\parallel} \sin(\delta_{\perp} - \delta_{\parallel} + \theta^c_{\perp} - \theta^c_{\parallel} ) \end{array}$
& $\begin{array}{c} -\lambda_{\perp} \cos(\delta_{\perp} - \delta_{\parallel} + \theta^c_{\perp} - \theta_{\parallel} ) \\ -\lambda_{\parallel} \cos(\delta_{\perp} - \delta_{\parallel} + \theta_{\perp} -\theta^c_{\parallel} ) \end{array}$
& $-2\sin^2\theta_1\sin^2\theta_2\sin 2\Phi$ \\ \hline
5       &$ {|A_\parallel||A_0|}/{2} $ 
& $\begin{array}{c} \cos(\delta_0 - \delta_{\parallel} + \theta_{0} - \theta_{\parallel} )  \\ +\lambda_{0} \lambda_{\parallel} \cos(\delta_0 - \delta_{\parallel} + \theta^{c}_{0} - \theta^c_{\parallel} )  \end{array}$
& $ \begin{array}{c} -\lambda_{0} \cos(\delta_0 - \delta_{\parallel} + \theta^c_0 - \theta_{\parallel} ) \\  -\lambda_{\parallel} \cos(\delta_0 - \delta_{\parallel} + \theta_0 - \theta^c_{\parallel} ) \end{array}$
& $\begin{array}{c} \cos(\delta_0 - \delta_{\parallel} + \theta_0 -\theta_{\parallel} )  \\ -\lambda_{0} \lambda_{\parallel} \cos(\delta_0 - \delta_{\parallel} + \theta^c_0 - \theta^c_{\parallel} ) \end{array}$
& $\begin{array}{c} -\lambda_{0} \sin(\delta_0 - \delta_{\parallel} + \theta^c_0 - \theta_{\parallel} ) \\ + \lambda_{\parallel} \sin(\delta_0 - \delta_{\parallel} + \theta_0 - \theta^c_{\parallel} ) \end{array}$
&$ \sqrt{2}\sin2\theta_1\sin2\theta_2\cos\Phi$ \\ \hline
 6       &$ {|A_0||A_\perp |}/{2} $          
& $\begin{array}{c} \sin(\delta_{\perp} - \delta_{0} + \theta_{\perp} -\theta_{0} )  \\ -\lambda_{\perp} \lambda_{0} \sin(\delta_{\perp} - \delta_{0} + \theta^c_{\perp} - \theta^c_{0} )  \end{array}$
& $\begin{array}{c} \lambda_{\perp} \sin(\delta_{\perp} - \delta_{0} + \theta^c_{\perp} - \theta_{0} ) \\  - \lambda_{0} \sin(\delta_{\perp} - \delta_{0} + \theta_{\perp} - \theta^c_{0} ) \end{array}$
& $\begin{array}{c} \sin(\delta_{\perp} - \delta_{0} + \theta_{\perp} - \theta_{0})  \\ +\lambda_{\perp} \lambda_{0} \sin(\delta_{\perp} - \delta_{0} + \theta^c_{\perp} - \theta^c_{0} ) \end{array}$
& $\begin{array}{c} -\lambda_{\perp} \cos(\delta_{\perp} - \delta_{0} + \theta^c_{\perp} - \theta_{0} ) \\ -\lambda_{0} \cos(\delta_{\perp} - \delta_{0} + \theta_{\perp} -\theta^c_{0} ) \end{array}$
& $-\sqrt{2}\sin2\theta_1\sin2\theta_2\sin\Phi $\\\hline
\end{tabular}
}
\caption{Coefficients of the time-dependent terms and angular functions used in Eq.~\eqref{timedep}. Amplitudes are defined at $t=0$.}
\label{timedeptablestrongCP}
\end{table}
\section{New physics model}\label{sec:npmodel}
The model we choose to use in our study is that of the \emph{Chromomagnetic dipole operator} \emph{$O_{8g}$}, which, for $\bar{b}\rightarrow \bar{s}g$ process, is given as follows:
\begin{align}\label{chromomagneticoperator}
O_{8g} = \frac{g_s}{8\pi^2} m_b \bar{b}_{\alpha} \sigma^{\mu\nu} (1 + \gamma^5) \frac{\lambda^a_{\alpha\beta}}{2}s_{\beta}G^a_{\mu\nu}.
\end{align}
Though Chromomagnetic operator is a SM operator, it is suppressed by $b$-quark mass $m_b$ (and its chirally-flipped counterpart is suppressed by $s$-quark mass $m_s$). However, it is very sensitive to several NP models, like the Left-Right symmetric class of models or SUSY, where it can undergo \emph{chiral enhancement} to overcome the quark mass suppression. 
The effective Hamiltonian for $\Delta B = 1$ decay including the Chromomagnetic operator is thus given by $(q \in \{d,s\})$
\begin{align}\label{effhamiltonian}
\Ham_{\rm{eff}} =-\frac{G_F} {\sqrt{2}} V^*_{tb}V_{tq}  
 \left[ \sum_{i=3}^{6} (C_i^{\rm{SM}} O_i) + C_{8g} O_{8g} + \tilde{C}_{8g} \tilde{O}_{8g}  \right] + \rm{h.c.}
\end{align}
The operator with tilde is obtained by changing the sign of $\gamma_5$ term in the definition of $O_{8g}$ to obtain the chirally-flipped counterpart.
\par 
\cng{In $B_s^0 \rightarrow \phi \phi$ decay, the final state can be split into three helicity/transversity states, giving us access to three helicity/transversity amplitudes, which make up the total amplitude.
However, \emph{the contribution from the Chromomagnetic operator is suppressed in transverse penguin amplitudes} (originally pointed out in \cite{kagan}, and verified by pQCD approach in \cite{bvv8}). Therefore, the NP contributions manifesting via Chromomagnetic operator should predominantly contribute to longitudinal amplitude. Therefore, the total longitudinal transversity amplitude can be written as}
\begin{align}\label{transamptotal}
\Amp^{\rm Total}_{0,\phi\phi} &= -\frac{G_F}{\sqrt{2}} V_{tb}^* V_{ts}  (\xi^{\rm SM}_{0}\mx_0^{\rm SM} + \xi^{\rm L}_{0}\mx_0^{\rm NP} - \xi^{\rm R}_{0}\mx_0^{\rm NP} ) ,
\end{align}
where $\mx^{\rm SM}$ and $\mx^{\rm NP}$ contains the contribution from the matrix elements for SM and NP operators, respectively. The $\xi^p_k$ ($k=\{0,\parallel,\perp\}$ and $p \in \{\rm{SM},\rm{L},\rm{R}\}$) are combinations of Wilson coefficients, and contain the weak phases (L and R represents left- and right-handed currents, coming from $O_{8g}$ and $\tilde{O}_{8g}$, respectively). The actual form of $\xi$ and $\mx$ depend upon the model chosen to compute the matrix elements, but it is not important for our purposes. Notice the \emph{sign change in the longitudinal component of right-handed amplitude}: this sign change occurs due to the sign change in the axial part of the current; we have verified this for longitudinal amplitude by both naive factorisation and pQCD approach.
\par 
Let us now see the observables which are sensitive to our NP model. 
Recalling the definition of interference phase from Eq.~\eqref{intphase} and putting Eq.~\eqref{transamptotal} in it, we can write 
\begin{align}\label{rationalisedphiphi}
 \frac{q}{p} \frac{\bar{\Amp}^{\rm Total}_{0,\phi\phi}}{\Amp^{\rm Total}_{0,\phi\phi}}  = \lambda_0 e^{-i(\theta_0 - \theta_0^c)} 
\end{align}
Since NP is contributing only to longitudinal amplitude, \emph{only $\lambda_0$ and $\theta_0-\theta_0^c$ would get contributions from NP, while CP-violating parameters of other transversities  would assume their SM values, i.e $\theta_\parallel = \theta_\parallel^c  = \theta_\perp = \theta_\perp^c =0$ and $\lambda_\parallel = \lambda_\perp = 1$. An observation of non-zero value of $\lambda_0-1$ and/or $\theta_0 - \theta_0^c$ would clearly indicate the presence of NP.}
\par
Let us now compare the phase scheme that LHCb used in their fit to ours.  Before comparing with our parametrisation, we note that the CP-violating phase in LHCb is defined as $\phi^{\rm LHCb}_k \equiv \theta_k - \theta^c_k$. 
LHCb uses the following two different fit configurations:
\begin{itemize}
\item LHCb {\it helicity-dependent} (HD) scheme: \\
$\phi^{\rm LHCb}_{0} = 0$, $\lambda_k = 1$ $\forall k$ ($\phi^{\rm LHCb}_{\perp}$ and $\phi^{\rm LHCb}_{\parallel}$ are the {CP-violating fit} parameters). 
\item LHCb {\it helicity-independent} (HI) scheme: \\
 $\phi=\phi_{k}^{\rm LHCb}$ $\forall k$, $\lambda = \lambda_k$ $\forall k$ ($\phi$ and $\lambda$ are the  {CP-violating fit} parameters). 
\end{itemize}
The new fit configuration we are proposing is
\begin{itemize}
\item  NP manifested via Chromomagnetic operator: \\
$\phi_\perp^{\rm LHCb}$=$\phi_\parallel^{\rm LHCb}$=0 or equivalently $\theta_\parallel = \theta_\parallel^c  = \theta_\perp = \theta_\perp^c =0$, $\lambda_\perp$=$\lambda_\parallel$=1
\\ ($\phi_0^{\rm LHCb}$ and $\lambda_0$ are the  {CP-violating fit} parameters). 
\end{itemize}
The LHCb fit configuration does not match to ours, and a new fit of LHCb data with this new scheme based on our model would be very interesting. We emphasise that neither of the two LHCb schemes above fit $\phi_0^{\rm LHCb}$ and $\lambda_0$ simultaneously; therefore, our phase scheme is a new avenue to search for NP manifesting itself via Chromomagnetic operator. 
\par

\section{Sensitivity study with the new fit configuration} \label{sec:sensitivitystudy}
Using the proposed fit scheme based on Chromomagnetic operator, we perform a sensitivity study on the CP-violating parameters.
To illustrate the fit, we first construct two {\it pseudo datasets} by using the LHCb best-fit values, denoted as Data HI and Data HD for the LHCb helicity-independent and helicity-dependent fit, respectively. 
Our fit results are shown in Table~\ref{table:dcp}.
We note that the results using Data HI and Data HD agree relatively well. 
The obtained uncertainty of $\sigma(\lambda_0)=6-7$\% and $\sigma(\theta_0-\theta_0^c)=5-6$\% with the currently available LHCb statistics (5 $\rm fb^{-1}$)  may be used as an indication for future studies. 
\begin{table}[ht]
\begin{center}
\begin{tabular}{|c|c|c|c|c|}
\hline
& \multicolumn{2}{c|}{Data HD} & \multicolumn{2}{c|}{Data HI} \\
\hline
Fit Parameter & Central Value & $\sigma$ & Central Value & $\sigma$ \\
\hline
$\lambda_0$ & 0.978 & 0.058 & 0.984 & 0.070 \\
\hline
$|A_0|^2$ & 0.386 & 0.025 & 0.385 & 0.032 \\
\hline
$|A_{\perp}|^2$ & 0.287 & 0.018 & 0.288 & 0.036 \\
\hline
$\theta_0 - \theta^c_0$ & -0.002 & 0.055 & 0.066 & 0.053 \\
\hline
$\delta_{\parallel} - \delta_{\perp}$ & -0.259 & 0.054 & -0.261 & 0.056 \\
\hline
$\delta_{\parallel} -\delta_{0} - \theta_0$ & 2.560 & 0.071 & 2.589 & 0.079 \\
\hline
\end{tabular}
\caption{Fit results based on our model assumptions, i.e. longitudinal component dominance for NP contributions coming from Chromomagnetic operator ($\theta^c_{\parallel} = \theta_{\parallel} = \theta^c_{\perp} = \theta_{\perp} = 0$ and $\lambda _{\parallel} = \lambda_{\perp} = 1$).}
\label{table:dcp}
\end{center}
\end{table}

\section{Summary}
In this work, we investigate a new physics search with the CP violation measurement of the $B_s^0 \rightarrow \phi\phi$ decay. In LHCb analysis, two types of NP scenarios were investigated, namely helicity-dependent and helicity-dependent. Here, we propose a new scenario based on the NP model induced by left- and right-handed Chromomagnetic operators, producing a new quark level $b\rightarrow s\bar{s}s$ diagram with an extra source of CP violation.  We do a sensitivity study of the CP-violating parameters of our proposed model, and we find that with the current statistics, they can be determined at $5-7\%$ precision.
\bibliographystyle{plainurl}
\bibliography{Bibliography.bib}

\begin{thebibliography}{1}

\bibitem{rosner}
M.~Gronau and J.~L. Rosner.
\newblock {Triple product asymmetries in $K$, $D_{(s)}$ and $B_{(s)}$ decays}.
\newblock {\em Phys. Rev. D}, 84:096013, 2011.
\newblock \href {http://arxiv.org/abs/1107.1232} {\path{arXiv:1107.1232}},
  \href {https://doi.org/10.1103/PhysRevD.84.096013}
  {\path{doi:10.1103/PhysRevD.84.096013}}.

\bibitem{kagan}
A.~L. Kagan.
\newblock {Polarization in $B \rightarrow VV$ decays}.
\newblock {\em Phys. Lett. B}, 601:151--163, 2004.
\newblock \href {http://arxiv.org/abs/hep-ph/0405134}
  {\path{arXiv:hep-ph/0405134}}, \href
  {https://doi.org/10.1016/j.physletb.2004.09.030}
  {\path{doi:10.1016/j.physletb.2004.09.030}}.

\bibitem{mixingcpasymmetry}
LHCb, R.~Aaij, et~al.
\newblock {Measurement of the $CP$ asymmetry in $B_s^0-\bar{B}_s^0$ mixing}.
\newblock {\em Phys. Rev. Lett.}, 117(6):061803, 2016.
\newblock [Addendum: Phys.Rev.Lett. 118, 129903 (2017)].
\newblock \href {http://arxiv.org/abs/1605.09768} {\path{arXiv:1605.09768}},
  \href {https://doi.org/10.1103/PhysRevLett.117.061803}
  {\path{doi:10.1103/PhysRevLett.117.061803}}.

\bibitem{lhcb}
LHCb, R.~Aaij, et~al.
\newblock {Measurement of CP violation in the $ {B}_s^0 \rightarrow \phi \phi$
  decay and search for the $ B^0\rightarrow \phi\phi $ decay}.
\newblock {\em JHEP}, 12:155, 2019.
\newblock \href {http://arxiv.org/abs/1907.10003} {\path{arXiv:1907.10003}},
  \href {https://doi.org/10.1007/JHEP12(2019)155}
  {\path{doi:10.1007/JHEP12(2019)155}}.

\bibitem{lhcbjpsimixing}
LHCb, R.~Aaij, et~al.
\newblock {Updated measurement of time-dependent CP-violating observables in
  $B^{0}_{s}\rightarrow J/\psi K^+ K^-$ decays}.
\newblock {\em Eur. Phys. J. C}, 79(8):706, 2019.
\newblock [Erratum: Eur.Phys.J.C 80, 601 (2020)].
\newblock \href {http://arxiv.org/abs/1906.08356} {\path{arXiv:1906.08356}},
  \href {https://doi.org/10.1140/epjc/s10052-019-7159-8}
  {\path{doi:10.1140/epjc/s10052-019-7159-8}}.

\bibitem{bvv8}
Da-Cheng Yan, Xin Liu, and Zhen-Jun Xiao.
\newblock {Anatomy of $B_s \rightarrow VV$ decays and effects of
  next-to-leading order contributions in the perturbative QCD factorization
  approach}.
\newblock {\em Nucl. Phys. B}, 935:17--39, 2018.
\newblock \href {http://arxiv.org/abs/1807.00606} {\path{arXiv:1807.00606}},
  \href {https://doi.org/10.1016/j.nuclphysb.2018.08.002}
  {\path{doi:10.1016/j.nuclphysb.2018.08.002}}.

\end{thebibliography}
\end{document}